\documentclass[sigconf,9pt]{acmart}
\usepackage{epsfig,subfigure,multirow}
\usepackage{tabularx}
\usepackage{enumitem}

\copyrightyear{2020}
\acmYear{2020}
\setcopyright{usgovmixed}\acmConference[WiseML '20]{2nd ACM Workshop on Wireless Security and Machine Learning}{July 13, 2020}{Linz (Virtual Event), Austria}
\acmBooktitle{2nd ACM Workshop on Wireless Security and Machine Learning (WiseML '20), July 13, 2020, Linz (Virtual Event), Austria}
\acmPrice{15.00}
\acmDOI{10.1145/3395352.3402619}
\acmISBN{978-1-4503-8007-2/20/07}

\begin{document}

\title{Adversarial Machine Learning based Partial-model Attack in IoT}

\author{Zhengping Luo}
\affiliation{%
University of South Florida.\\
Email: zhengpingluo@usf.edu.
}

\author{Shangqing Zhao}
\affiliation{%
University of South Florida.\\
Email: shangqingzhao@usf.edu.
}

\author{Zhuo Lu}
\affiliation{%
University of South Florida.\\
Email: zhuolu@usf.edu.
}

\author{Yalin E. Sagduyu}
\affiliation{%
Intelligent Automation Inc.\\
Email: ysgduyu@i-a-i.com.
}
\author{Jie Xu}
\affiliation{%
University of Miami.\\
Email: jiexu@miami.edu.
}

\email{}

\newcommand{\rev}[1]{{\color{red}#1}} 

\begin{abstract}
As Internet of Things (IoT) has emerged as the next logical stage of the Internet, it has become imperative to understand the vulnerabilities of the IoT systems when supporting diverse applications. Because machine learning has been applied in many IoT systems, the security implications of machine learning need to be studied following an adversarial machine learning approach. In this paper, we propose an adversarial machine learning based partial-model attack in the data fusion/aggregation process of IoT by only controlling a small part of the sensing devices. Our numerical results demonstrate the feasibility of this attack to disrupt the decision making in data fusion with limited control of IoT devices, e.g., the attack success rate reaches 83\% when the adversary tampers with only 8 out of 20 IoT devices. These results show that the machine learning engine of IoT system is highly vulnerable to attacks even when the adversary manipulates a small portion of IoT devices, and the outcome of these attacks severely disrupts IoT system operations.
\end{abstract}

\ccsdesc[500]{Security and privacy~Mobile and wireless security}
\ccsdesc[500]{Networks~Network reliability}

\keywords{Internet of Things, wireless security, machine learning, adversarial machine learning, data fusion}

\maketitle

\section{Introduction}
\emph{Internet of Things} (IoT) is cast as a system of networked devices embedded with sensors to gather and interchange data, and execute complex tasks \cite{atzori2010internet,gubbi2013internet,khan2020industrial}. As technology is advancing from web2 (social networking web) to web3 (ubiquitous computing web), IoT, as an extension of Internet into the physical world, becomes the core technology to connect sensors and actuators into an integrated network \cite{fernandes2016security,zanella2014internet, celik2019program}. Applications of IoT include but are not limited to smart home, smart warehouse, vehicular networks, environmental monitoring, and perimeter security \cite{khan2020industrial}.

\emph{Wireless Sensor Network} (WSN) is often considered as the building block for the IoT systems. However, the sensing devices in IoT are prone to failures \cite{gubbi2013internet}. While information exchange among heterogeneous sensing devices/actuators and hubs/data centers is mandatory, many applications of IoT have strict timing, security, reliability requirements. Therefore, how to ensure the real information sensed by sensors/actuators to be securely received by the hub/data center in a wireless environment is critical for both the security and reliability of the IoT systems.

As the scale of IoT systems grows rapidly with more devices added, \emph{machine learning} has started playing a key role in the processing and learning from large-scale data generated by IoT devices \cite{mahdavinejad2018machine}. While machine learning helps with efficient operation of IoT systems, the other side of the coin is concerning adversaries may also employ machine learning as powerful means to launch attacks against IoT infrastructures. The study of machine learning under adversaries is referred to as \emph{adversarial machine learning} \cite{Vorobeychik1}.

In this paper, we focus on the security of \emph{data fusion/aggregation} process in IoT. Original data or information is collected through IoT devices such as actuators, RFID, switches, and sensors. Then multiple IoT devices report their data to a hub or data center to aggregate and report the aggregated results to the cloud or data analysis center. In this paper, we consider a scenario where multiple devices report their data to a fusion center, and the fusion center makes a binary decision based on the received information. In this scenario, we show that the adversary can employ machine learning techniques to launch an attack by controlling only a small part of the devices, which we call the \emph{partial-model attack}.

The main contributions of this paper are listed as: (i) We introduce a machine learning based partial-model attack in IoT data fusion process, where the adversary aims to disrupt decision making of IoT data fusion process by taking advantage of the IoT device properties; (ii) We present numerical experiments to validate the proposed attack framework and demonstrate that the successful attack ratio is high even when a small portion of sensors are controlled by the adversary. For instance, in a scenario where 8 out of 20 devices are controlled by an adversary, the hit ratio reaches up to 83\%; (iii) We discuss potential ways of defending against machine learning based partial-model attack in IoT systems.


\section{Background and Related Work}\label{sec:background}

\subsection{IoT architecture}
IoT finds rich applications including but not limited to Industrial Internet of Things (IIoT), smart home, smart city, healthcare, and transportation. Basic components that enable the IoT benefits include: (i) \emph{hardware} (heterogeneous sensors and actuators); (ii) \emph{middleware} (data aggregation/fusion center and storage devices); (iii) \emph{presentation} (visualization and other analysis tools that enable access to different platforms) \cite{gubbi2013internet,barriga2016proposal,liu2013role}.



Radio Frequency Identification (RFID) is a major source of data for many IoT systems \cite{plageras2017solutions,stergiou2020secure}. It is a technology that employs electromagnetic fields for data transfer and automatic object detection. With the RFID tag, items can be detected by reading their labels. Once sensor data is collected, the next step is to transfer data for storage and processing. Cloud computing is the storage and computing center of IoT, where data analytics are based on. Users can access the cloud computing and generate visual presentation of the collected data \cite{stergiou2017recent}. Moreover, cloud platforms provide device lifecycle management for IoT and can provide digital twin version of real systems \cite{firouzi2020machine}.

\begin{figure}[htp]
\centering
\includegraphics[scale = 0.55,trim={0 0cm 0 0cm},clip]{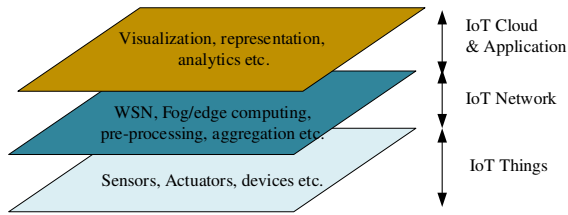}
\caption{The generalized layers of typical IoT systems.}
\label{fig:framework}
\end{figure}

We illustrate the general architecture of IoT systems in Fig \ref{fig:framework} as three layers: \emph{IoT things} (physical devices), \emph{IoT network}, and \emph{IoT cloud and application}. The bottom layer consists of the physical devices, the second layer focuses on the infrastructure such as network and data aggregation, the last (highest) layer is the user-oriented layer. 

\subsection{Adversarial machine learning and IoT}

Machine learning, especially \emph{deep learning}, has attracted tremendous attention since its successful application in image recognition \cite{krizhevsky2012imagenet}. Recently, deep learning has started finding applications also in wireless systems, including waveform design, signal analysis and security \cite{erpek1}. Devices within an IoT system often generate data continuously and simultaneously at high rates. To deal with this large-scale data, machine learning offers automated means to process and analyze data, and make decisions \cite{gubbi2013internet}. Unlike traditional statistical models, machine learning provides a way to learn parameters from the data, and it can make decisions based on both historical data and real time streaming data. Besides, given the heterogeneity of IoT systems, machine learning can be performed in either central or distributed fashion.

Despite its strengths, machine learning itself has many vulnerabilities that might be exploited by malicious users \cite{papernot2017practical,kurakin2016adversarial}. The security problem of IoT systems are critical for both the users and owners of IoT infrastructure. Machine learning in the presence of adversaries is studied under the emerging area of \emph{adversarial machine learning} \cite{kurakin2016adversarial, papernot2017practical, Vorobeychik1, Shi2017, Shi2018bc}. The shared nature of wireless medium makes machine learning especially vulnerable to various attacks built upon adversarial machine learning. In wireless domain, adversarial machine learning has been applied to launch different types of attacks \cite{Sagduyu2020}, including inference (exploratory) attack \cite{Shi2018, terpek}, evasion attack \cite{Kim, Kim2, Yalin2019, Sagduyu1}, poisoning (causative) attack \cite{Sagduyu1, YiMilcom2018, Luo2019}, Trojan attack \cite{Davaslioglu1}, and spoofing attack \cite{Shi}. These attacks are stealthier (more difficult to detect) and operate with lower footprint compared with conventional wireless attacks such as a jamming \cite{Sagduyu2008}. Adversarial learning can also be used to augment training data with synthetic data samples \cite{KemalICC}.

%

Security of IoT has drawn increasing attention. Many of the security studies of IoT are centered around two fronts \cite{celik2018sensitive}: sensing end-devices and connecting protocols. Strategies such as improving security through firewall and mobility policies have been presented in \cite{kubler2015standardized}. Intrusion detection mechanisms formulated as anomaly detection have been discussed for IoT systems in \cite{zarpelao2017survey}. The privacy issues of IoT have been considered in \cite{goad2020privacy}.

Major security vulnerabilities and challenges of IoT can be summarized as follows \cite{firouzi2020machine}:
\begin{itemize}
\item \textit{Sub-system heterogeneity}: Devices, sensors, actuators and sub-controlling components within IoT systems are heterogeneous. Thus, it is challenging to integrate them into one system, and security measures required for different subsystems might also differ from each other. It is necessary to find a common strategy to control all the heterogeneous sub-systems.
\item \textit{End-device reliability}: End-devices of IoT systems are distributed in real world and they can be influenced by various environmental factors that may cause them to fail to function, report wrong sensing results, or even lose control to malicious users. How to ensure the reliability of the end-devices is critical to the security of IoT.
\item \textit{Data security}: Data security in IoT systems involves multiple concerns such as safe transmission (how to guard the sensed information such that it can be safely transferred to the cloud/processing center) and safe operation of data center.
\end{itemize}

Our paper aims to employ adversarial machine learning techniques to launch attacks against the \emph{IoT data fusion process}. The adversary first learns a machine learning model and then based on the learned model it crafts adversarial inputs. Detailed information on this attack is given in the next section.

\section{Adversarial machine learning-based partial-model attack}\label{sec:attack}
\subsection{IoT data fusion}
Data gathering from multiple devices is a critical step in IoT systems. Compared with making decisions from a single data source, collecting data from multiple sensing devices may help filter out noises and other deviations \cite{stillman2001towards}. 
The requirements of IoT data fusion are summarized as follows \cite{ding2019survey}: (i) \emph{Context-aware}: It is necessary to support adaptive and flexible services. The context information like location, weather and other environmental factors may change. Thus, data fusion at IoT systems needs to calibrate and adapt to these changes. (ii) \emph{Privacy preserving}: It is necessary to protect privacy of sensitive IoT information such as personal habits or industrial secrets. (iii) \emph{Reliable}: It is necessary to detect, remove or replace unreliable devices, as the sensed information may be noisy or contaminated. (iv) \emph{Real time}: It is necessary to make timely decisions to support real-time operations of data fusion. (v) \emph{Verifiable}: It is necessary to keep the fusion result verifiable to the user or public.


\begin{figure}[htp]
\centering
\includegraphics[scale = 0.8,trim={0 0cm 0 0cm},clip]{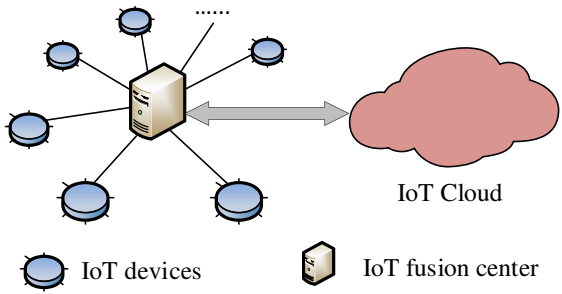}
\caption{The simplified sensing and decision model of IoT.}
\label{fig:fusion}
\end{figure}

In this paper, we consider a general scenario of IoT data fusion as shown in Fig.\ref{fig:fusion}. Multiple IoT devices report their collected information to a data fusion center, and the data fusion center makes a decision (classification or regression), and then transfers the decision output to the IoT cloud for further analysis.

\subsection{Partial-model attack}

\begin{figure*}[h]
\centering
\includegraphics[scale = 0.8,trim={0 0cm 0 0cm},clip]{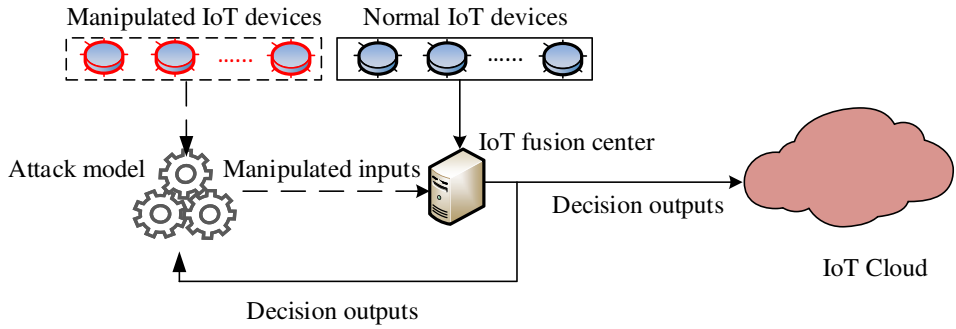}
\caption{The machine learning based partial-model attack in IoT.}
\label{fig:attack}
\end{figure*}

The sensing devices play a key role in IoT systems. Therefore, secure and efficient communication between sensing devices and the data fusion center is needed to support IoT operations. However, sensing devices are prone to failures and manipulation by adversaries. We propose an adversarial machine learning based attack model as follows. We assume that the adversary controls a small number of IoT devices, and the adversary knows the decision output of the IoT fusion center. However, the adversary has no knowledge about the decision process in the IoT fusion center. Data is exchanged between the IoT devices fusion center and other edge computing center over wireless channels. Thus, it is possible for the adversary to hijack the over-the-air transferred information.

We assume that there are $n$ IoT sensing devices that report their information to the IoT data fusion center to aggregate and output a decision. In the meantime, $m$ of these devices are controlled by a malicious adversary. By controlling a small portion of IoT devices based on machine learning techniques, the adversary aims to take advantage of the failures of other normal/un-manipulated sensing nodes, further flip or change the decision output significantly. One application is spectrum sensing data falsification by some rogue nodes in cooperative spectrum sensing \cite{SSDF, Luo2019}. As IoT devices may fail or report confused information, the adversary can detect this kind of uncertainty and further take advantage of this uncertainty to expand the impact of the attack.

For a successful attack, the adversary first needs to learn about the potential true decision/classification output in the IoT fusion center from controlled IoT sensing devices and historical decision/classification output. Therefore, the adversary needs to build first a machine learning model to infer the potential decision state based on the information collected from controlled IoT sensing devices. For the attack model, the inputs come from the sensing results of controlled devices. The output are the adversarial vectors crafted from the inputs.

\begin{figure}[htp]
\centering
\includegraphics[scale = 0.6,trim={0 0cm 0 0cm},clip]{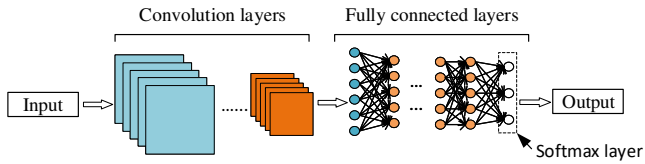}
\caption{A typical convolutional neural network.}
\label{fig:dnn}
\end{figure}

The next question is when to launch the attack. In our proposed partial-model attack, the adversary does not launch the attack at each round of inputs. The attack should be launched when controlled devices sensed possible confused signals, i.e., when the learned decision model by the adversary is less certain about the decision. Consider a convolutional neural network (shown in Fig.\ref{fig:dnn}), in which feature vectors are served as inputs of the framework. The next layers are the convolutional layers, which consist of convolutional and pooling operations. The main objective of the convolutional layers is to extract more complicated features (e.g., silhouettes in image recognition). The fully connected layers follow the convolutional layers and aim to find the optimal combination of the previous features. The last layer of the framework is named as SoftMax layer. The number of neurons in the SoftMax layer is equal to the output classes.

The value in SoftMax layer is considered as ``confidence value'' and reflects how ``confident'' the trained model is towards the output \cite{fredrikson2015model}. Therefore, we employ the largest value of the neuron output in the SoftMax layer, which is also the value of the final decision output for the model, as an indication about the certainty of des. The output value of the SoftMax layer is the result of a squashing function, which limits the output within the range between 0 and 1. Mathematically the standard SoftMax function is defined as: $\sigma(z_i) = \frac{e^{z_i}}{\sum_{j=1}^{c}e^{z_j}},$ in which $\sigma(z_i)$ is the output, or confidence value of the final decision towards the $i_{th}$ class. $c$ is the total number of output classes. Each output value in the SoftMax layer gives the ``confidence'' of the decision output towards each class \cite{fredrikson2015model}. When the confidence value is beyond a certain threshold, malicious inputs are generated and sent to the IoT data fusion center, otherwise normal data are sent to avoid being detected.

After the learning step, the adversary infers the potential true decision output. Then through manipulating the information of controlled IoT devices sent to the IoT data fusion center, the adversary has the possibility to compromise the IoT data fusion center. There are different ways to craft malicious inputs (a.k.a. adversarial inputs) for the controlled IoT sensing devices as shown in\cite{kurakin2016adversarial,papernot2017practical}. The basic idea of crafting adversarial inputs is to move the input towards the decision boundary of the learned classification model such that the modification is minimized.

The overall attack framework is shown in Fig.\ref{fig:attack}. Normal IoT devices report the collected data directly to the IoT fusion center, while manipulated IoT devices need to report the collected information to the adversary. The adversary first learns an attack model and then in later rounds decides whether to launch attack, or not. When the adversary launches the attack, it reports the manipulated inputs to the IoT fusion center, otherwise it reports the original data. The proposed partial-model attack is not a mathematically guaranteed attack. The key for this attack to succeed depends on the overall uncertainty among IoT devices. In the next section, we present simulation results to evaluate the proposed attack.

\section{Experimentation and analysis}\label{sec:experiment}
We conduct detailed simulations of the IoT data fusion process and analyze the performance of the partial-model attack in this section.

\subsection{Experimental configurations}
The IoT fusion center in our attack model collects information from a set of sensing devices, aggregates them to make a decision and delivers the decision to the IoT cloud for further data analysis. In our simulation, the decision model in the IoT data fusion center is set as a binary classification model as many sensing tasks are binary, such as switches and signal sensing devices. We assume that there are 20 IoT sensing devices such as RFID. 10000 data samples are collected from two Gaussian distributions, each of the distribution corresponds to one class. The first 2000 data samples serve as training data for the adversary. The remaining data samples are used for evaluation. To make the simulation consistent with real environment uncertainties, the mean and deviation of the Gaussian distribution for each device are set as a random number within a given range (e.g., to represent the potential differences in oscillators when spectrum data is sensed). The adversary employs a 5-layer neural network as the learning model. The implementation of the learning model is based on TensorFlow.

In the IoT data fusion center, we employ Support Vector Machine (SVM) as the fusion rule, which is one of the most popular statistical classification models. It is worth noting that other fusion rules such as multi-layer perceptron neural network, decision tree, etc. can also be used as the fusion rule. Due to the limitation of space, we consider SVM in this paper.

\subsection{Attack performance analysis}
The performance of the proposed machine learning based partial-model attack is measured by hit ratio (namely, attack success ratio), which is defined as:
\begin{equation*}
\text{Hit ratio} = \frac{\text{The number of successful attacks}}{\text{Total number of attack instances}}.
\end{equation*}
The successful attack here corresponds to those input samples that successfully flip the decision of IoT fusion center that would have been made when no attack is launched.

\begin{figure}[h]
\centering
\includegraphics[scale = 0.4,trim={0 0cm 0 0cm},clip]{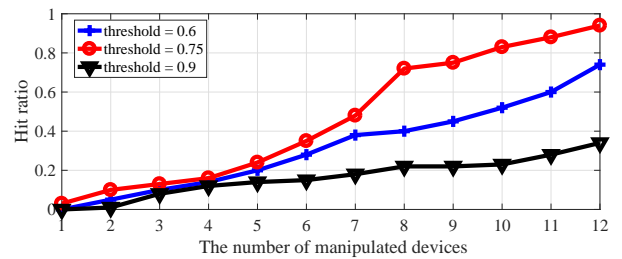}
\caption{The relationship of the number of controlled sensing devices with the hit ratio.}
\label{fig:exper}
\end{figure}

We first evaluate the hit ratio by varying the number of controlled devices $m$. The results under different confidence threshold value of 0.60,0.75 and 0.9, respectively, are shown in Fig.\ref{fig:exper}. The hit ratio increases with $m$. In particular, when $m$ is 8, which is less than half of the total number of devices, $n$, the hit ratio is 72\% while the confidence threshold value is 0.75. As $m$ further increases, the hit ratio approaches to 1. When the threshold is too large, the number of attacks will decrease dramatically, thus the overall hit ratio will also decrease.

\begin{figure}[h]
\centering
\includegraphics[scale = 0.45,trim={0 0cm 0 0cm},clip]{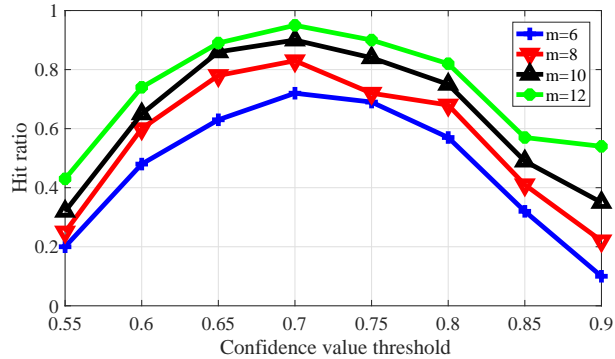}
\caption{Hit ratios under different confidence thresholds and different number of manipulated devices.}
\label{fig:confidence}
\end{figure}

Next, we evaluate the relationship between the hit ratio and the confidence threshold. The results are shown in Table \ref{fig:confidence}. Four different scenarios when $m$ is 6, 8, 10, or 12 are considered. We observe that when the confidence threshold is set to near 0.5 (the confidence threshold is always larger than 0.5 due to our binary model), the attack success ratio is comparatively lower than the case when the threshold is set around 0.7. The reason is that when the learned model is less certain about the decision output, it has higher probability that the learned model makes mistakes in inferring the potential true decision in the IoT data fusion center.

When the threshold is set as 0.7, the hit ratio approaches to 83\% when $m=8$. The hit ratio decreases dramatically when the confidence value increases beyond 0.8. The reason is that when the threshold is set too high, the number of attacks increases and thus it becomes difficult for the adversary to take advantage of the uncertainty of other normal IoT sensing devices.

Our simulation results demonstrate that the partial-model attack is likely to succeed when the IoT devices or information exchange involve uncertainties. The adversary takes advantage of the ``uncertainty'' of other normal devices and increases its hit ratio. Therefore, robust security mechanisms are needed in future IoT system design.

\section{Discussion}\label{sec:discussion}
As machine learning provides IoT systems with powerful means of learning from data and solving complex tasks, it also raises security concerns due to its vulnerability to adversarial manipulation. Our proposed adversarial machine learning based partial-model attack model focuses on the IoT data fusion process and equips the adversary with the capability to launch successful attacks even when the adversary controls a small part of the IoT devices by exploiting the performance uncertainty of the IoT devices or the communication channel. How to counter the proposed attack is our future work. Below, we provide several potential mechanisms for defense: \textit{Deploying robust anomaly detection mechanism in the IoT fusion center}. This is a direct method to defend the IoT systems against the partial-model attack. However, in this attack, all the manipulated devices cooperate with each other to launch the attack. Thus, how to design an anomaly detection method to detect a set of devices is a challenge. \textit{Improving privacy protection in every level of the IoT infrastructure}. In the partial-model attack, the key to learn a partial model to mimic the fusion center is the availability of the output of the fusion center. Thus, the decision information can be kept as private and secure by deploying a privacy protection mechanism.

Using machine learning to attack the IoT systems is detrimental to the IoT security. On the other hand, machine learning can be also employed as a defense method \cite{doshi2018machine}. Therefore, it is important to understand the interaction of machine learning techniques used for attack and defense, and game theory can be used as mathematical means to study the conflict of interest driven by machine learning.

\section{Conclusion}\label{sec:conclusion}
How to protect the privacy and security of IoT systems from the data collecting stage to the final visualization and application stage is essential to the successful adoption of IoT. In this paper, we introduced an adversarial machine learning based partial-model attack strategy, which mainly sits in the data collecting and aggregating stage of IoT systems. We use the machine learning based model to infer the potential decisions or aggregate results and then launch attacks by manipulating the data of the controlled IoT devices. Simulations show that the attack is highly successful even with a small part of manipulated IoT devices.
~\\

\noindent{\bf Acknowledgement:} The work at USF was supported in part by NSF CNS-1717969.



\bibliographystyle{ACM-Reference-Format}
\bibliography{sample-base}

\end{document}